\documentclass[12pt]{iopart}

\usepackage{iopams}
\usepackage{amssymb}
\usepackage{setstack}
\usepackage{graphicx}% Include figure files
\usepackage{dcolumn}% Align table columns on decimal point
\usepackage{bm}
\usepackage{dsfont}
\usepackage{amssymb}

\newcommand{\atanh  }{{\rm{atanh}}}

\newcommand{\sech   }{\mathrm{sech}}

\usepackage{color}

\begin{document}
\date{\today}
\title{\bf The phase diagram of L\'evy spin glasses}
\author{I. Neri, F. L. Metz and D. Boll\'e}
\begin{abstract}
We study the L\'evy spin-glass model with the replica and the cavity method.  In this model each spin interacts through a finite number of strong bonds and an infinite number of weak bonds.   This hybrid behaviour of L\'evy spin glasses becomes transparent in our solution: the local field contains a part propagating along a backbone of strong bonds and a Gaussian noise term due to weak bonds.  Our method allows to determine the complete replica symmetric phase diagram, the replica symmetry breaking line and the entropy.  The results are compared with simulations and previous calculations using a Gaussian ansatz for the distribution of fields.  
 \end{abstract}

\section{Introduction}

The prototype mean-field model of spin glasses
is the Sherrington-Kirkpatrick (SK) model \cite{SH1975,SK1978,par1987}. 
In this fully-connected (FC) system any pair of spins is coupled through 
weak interactions, of order $\mathcal{O}(N^{-1/2})$ in the total number 
of spins $N$, whose values are drawn independently from a 
Gaussian distribution.

Within the assumption that the free-energy landscape contains one single valley, the 
effective field on a given spin
is a sum of a large number of uncorrelated random 
variables with a finite variance and
%variables
%arising from the interactions with neighbouring spins, 
the usual central limit theorem (CLT) holds. As a consequence, the effective field
follows a Gaussian distribution fully characterized by its 
first two moments, leading to a description in terms of two observables:
the magnetization and the Edwards-Anderson order parameter.
The CLT reflects the
independence of the macroscopic behavior of the system
with respect to the details of the coupling
distribution. The existence of a CLT in the SK model
is technically very convenient 
and simplifies the replica and the cavity
method at high temperatures $T$. At low temperatures extreme values become important and a more complicated description is necessary. The success of the cavity and replica method lies in the detailed and exact description they give 
of the intricate behavior of the SK model at low temperatures, characterized
by the presence of several degenerate states separated
by infinite barriers \cite{par1987}.

However, it was shown that materials composed of magnetic 
impurities, randomly distributed in a non-magnetic host and 
interacting through the RKKY dipolar potential, exhibit a Cauchy distribution
of effective fields.  This is in particular true for a small concentration of magnetic
impurities \cite{M68,KHZ76}. An analogous
result was obtained in a spatially disordered system of
particles with dipolar interactions \cite{B96}. 
These results suggest that the choice of a 
coupling distribution that allows a wider variation of
coupling strengths  would be more realistic than 
the traditional Gaussian assumption used in mean-field models for spin glasses.

Disordered systems in which the randomness of the disorder variable $J$ is modelled by
a distribution $P(J)$ that has a power-law decay 
$P(J) \sim |J|^{-1-\alpha}$ ($\alpha < 2$), for large $|J|$, have
attracted less interest. A possible reason is the technical challenge to deal with distributions that do not fulfill the classical CLT.  The
heavy tails of $P(J)$ give rise to the divergence of
the second moment of the distribution, which invalidates the application
of classical CLT. In this case, the
generalized CLT of L\'evy and Gnedenko
holds  \cite{L1937, Gn1954}, and the sum of a large number
of independent random variables drawn from $P(J)$ 
follows the same distribution as the individual
summands, exhibiting only different scale factors.
The role of the large tails of $J$  
has proven to be crucial to the
long time or large size properties of
different disordered systems \cite{BBP07}.
As examples in
this context, we mention the theory of random matrices \cite{CB94,Burda2006,Bouchaud2007}, 
diffusion processes \cite{BG90} and the
portfolio optimization problem in theoretical finance \cite{GBP98}.

A FC model of spin glasses with interactions 
drawn from a distribution with power-law tails (a L\'evy spin glass)
was introduced
by Cizeau and Bouchaud \cite{Ciz1993}.   In L\'evy spin glasses 
every spin interacts with infinitely many weak bonds of order 
$\mathcal{O}\left(N^{-1/\alpha}\right)$ and a finite number of strong 
bonds of order $\mathcal{O}\left(1\right)$. In this sense the model is a hybrid 
between a FC spin glass, like the SK model, and a finitely connected (FiC) spin glass, like the Viana-Bray model \cite{viana1985}. 
The authors of \cite{Ciz1993} studied the model with the cavity method under the assumption that the distribution of effective fields is Gaussian. They found a spin glass phase stable under replica symmetry breaking and it was conjectured that at zero temperature the stability of replica symmetry is restored for $\alpha<1$.  
Recently, this model has
been studied with replica theory \cite{Hartm2008}. 
The effective field distribution is not Gaussian.  In \cite{Hartm2008} a complete phase diagram and a discussion of replica symmetry breaking was not given.

The purpose of this paper is to improve upon the foregoing studies by deriving the complete phase diagram without the Gaussian assumption, the entropy and the stability to replica symmetry breaking effects.  
We propose a method that 
consists in the insertion of a small cutoff in the 
distribution of the couplings $P(J)$, which
gives rise to a natural distinction between  
``weak bonds'' and ``strong bonds''.   
This allows us to solve the problem
through  both the replica method and the cavity method.  
We obtain a solvable self-consistent equation for the distribution
of effective fields.
Formally this equation is similar to the self-consistent equation for the effective field distribution of a FiC spin-glass system on a 
random graph \cite{MP87} and a straightforward implementation of the population 
dynamics algorithm \cite{Mez2003} is possible.
Therefore, the procedure allows us to obtain the complete phase
diagram of the model for all L\'evy distributions in contrast to previous works \cite{Ciz1993, Hartm2008}.
We include a 
skewness parameter in the definition of the model, 
responsible for controlling the
relative weight between the positive and
the negative tails of the coupling distribution.
The dependence of the different phases on this parameter is shown in the phase diagrams.  The results are compared with simulations.  
We calculate
the entropy of the system and the stability against replica symmetry breaking.   Our results
are compared with those obtained by Cizeau and Bouchaud \cite{Ciz1993}.

The paper is organized as follows. In section \ref{sec:def}, we
define the model.  We explain how to solve the model through replica theory in section \ref{sec:replica} and through the cavity method in section \ref{sec:cavity}.  In these sections we calculate the distribution of effective fields and compare them with the Gaussian assumption on the distribution of those fields.  The behavior of the magnetization is compared with simulations. In section \ref{sec:RSB} we derive the stability condition against replica symmetry breaking. 
The order parameter equations, derived in sections \ref{sec:replica}, \ref{sec:cavity} and \ref{sec:RSB}, are solved numerically to obtain the phase diagrams and the entropy in sections \ref{sec:phase} and \ref{sec:entropy}.   In section \ref{sec:conclusion} we present our conclusions.   The effect of the different parameters of the L\'evy distributions is shown in appendix A.  
Some details of the replica calculations are 
given in appendix B.

%%%%%%%%%%%%%%%%%%%%%%%%%%%%%%%%%%%%%%%%%%%%%%%

%%%%%%%%%%%%%%%%%%%%%%%%%%%%%%%%%%%%%%%%%%%%%%%

\section{The L\'evy spin glass} \label{sec:def}

We study a FC system
of $N$ Ising spins $\sigma_i = \pm 1$ ($i=1,\dots,N$)
with the Hamiltonian
\begin{equation}
H= - \sum_{i < j} J_{i j} \sigma_{i} \sigma_{j} , \label{hamilton}
\end{equation}
where the symmetric couplings $\{ J_{ij} \}$ are i.d.d.r.v.~drawn from a stable distribution $P^{J_1, \gamma, J_0 }_{\alpha}(J)$.
We define the stable distributions
$P^{J_1, \gamma, J_0 }_{\alpha}(J)$  through their characteristic function $L^{J_1, \gamma, J_0 }_{\alpha}(q)$: 
\begin{eqnarray}
 P^{J_1, \gamma, J_0 }_{\alpha}(J) \equiv \int \frac{dq}{2\pi}\exp\left(-iqJ\right)L^{J_1, \gamma, J_0 }_{\alpha}(q)
\label{defStable}.
\end{eqnarray}
The characterstic function is of the form
\begin{eqnarray}
\fl
 L^{J_1, \gamma, J_0 }_{\alpha}(q)  = \exp\left[ i\frac{q J_0 }{N} - \left|\frac{J_1q}{\sqrt{2} N^{1/\alpha}}\right|^{\alpha}
\left( 1-i \gamma \Phi {\rm sign}(q) \right) \right]
\label{defChar}. 
\end{eqnarray}
The distribution $P^{J_1, \gamma, J_{0} }_{\alpha}(J)$
is characterized by four parameters: the exponent 
$\alpha \in (0,1) \cup (1,2]$, the skewness 
$\gamma \in [-1,1]$, the scale parameter $J_1>0$ and 
the shift $J_0 \in \mathbb{R}$.   The quantity $\Phi$ is given by $\Phi=\tan{(\frac{\alpha \pi}{2})}$.  The scaling with $N$ in eq. (\ref{defChar}) ensures that the Hamiltonian (\ref{hamilton}) is of order $\mathcal{O}(N)$.  L\'evy distributions contain two different parameters that control the bias in the couplings: $J_0$ and $\gamma$.   We refer the reader to appendix A
for a discussion of the role of $\alpha$ and $\gamma$.    For $\alpha=1$ and $\gamma\neq0$ the quantity $\Phi$ has a different expression and we will not consider this case in the sequel.

The SK model is obtained
for $\alpha=2$ independent of $\gamma$: in this case the distribution $P^{J_1, \gamma, J_0 }_{\alpha}(J)$
is Gaussian with mean $J_0/N$ 
and variance $J_1^2/N$ \cite{SH1975}.
For $\alpha < 2$ and $-1 < \gamma < 1$, the asymptotic behavior $\rho(J)$
of $P^{J_1, \gamma, J_0 }_{\alpha}(J)$ for $|J| \rightarrow \infty$ can
be derived from the explicit form of $L^{J_1, \gamma, J_0 }_{\alpha}(q)$:
\begin{equation}
\rho(J)\equiv N\lim_{|J| \rightarrow \infty}  P^{J_1, \gamma,  J_0 }_{\alpha}(J) = 
\left( 1+\gamma {\rm sign} J \right) \frac{C_{\alpha}}{|J|^{\alpha+1}} ,
\label{asympP}
\end{equation}
where
\begin{equation}
C_{\alpha} = \left( \frac{J_1}{\sqrt{2}} \right)^{\alpha}\frac{1}{\pi} 
\sin\left(\frac{\alpha \pi}{2}\right)\Gamma(\alpha+1) . 
\end{equation}
Accordingly, the integrals for the second and higher moments of
the distribution diverge for $\alpha < 2$ due to the 
power-law decay illustrated by eq.~(\ref{asympP}).

We can define stable distributions through (\ref{defChar}) without losing any generality, see for example \cite{StochPaul, Nolan}. We 
remark that there are many equivalent definitions possible for the characteristic 
function $L$, see \cite{Nolan}. 
Loosely speaking, a random variable $x$ is stable if the
sum of a given number of independent and identical copies of $x$
is characterized by the same distribution as
the original variable, exhibiting only a different scale
and shift.

%%%%%%%%%%%%%%%%%%%%%%%%%%%%%%%%%%%%%%%%%%%%%%%%%%%%%%%%%%%%%%

\section{The replica method}\label{sec:replica}

\subsection{The distribution of effective fields}

In order to study the thermodynamic behavior of the L\'evy spin glass we employ the replica method \cite{par1987}. 
The partition function of the system at inverse temperature
$\beta= T^{-1}$ is defined by
\begin{equation}
Z = \sum_{\left\{\sigma\right\}_{i=1..N}} \exp{\Big[-\beta H\left(\left\{\sigma\right\}_{i=1..N}\right)\Big]} ,
\end{equation}
with $H\left(\left\{\sigma\right\}_{i=1..N}\right)$ given by eq.~(\ref{hamilton}). The averaged 
free-energy per spin $f$ can be written as follows
\begin{equation}
f = - \lim_{N \rightarrow \infty}\lim_{n \rightarrow 0} 
\frac{1}{\beta N n} \ln{\overline{Z^{n}}}. \label{frepl}
\end{equation}
The symbol $\overline{(\dots)}$ denotes the
average over the quenched random couplings 
$\{ J_{ij} \}$ with the distribution  $P^{J_1, \gamma, J_0}_{\alpha}(J)$.
The quantity $\overline{Z^{n}}$ is computed for positive
integers $n$ and the
limit $n \rightarrow 0$ is taken through an analytic continuation to real values.

However, the integer moments
$\overline{Z^{n}}$ of the partition function diverge 
for real $\beta$ due to the power-law behavior of 
$P^{J_1, \gamma, J_0 }_{\alpha}(J)$ for 
$|J| \rightarrow \infty$.
As noted in reference \cite{Hartm2008}, the
introduction of an imaginary temperature 
$\beta= - i k$, with a real parameter $k > 0$, allows
a straightforward calculation of the average $\overline{Z^{n}}$ 
by means of the definition of the characteristic function, 
eq.~(\ref{defChar}). However, it is
not possible to write the averaged $\overline{Z^{n}}$ 
in terms of the two
standard order-parameters usually employed in
the description of FC systems, i.e., the
magnetization and the spin-glass order parameter.
Therefore, it is necessary to use
the replica
method, as developed to deal with FiC spin glasses \cite{Monasson1998}.  
The macroscopic behavior is characterized in terms of a non-Gaussian effective
field distribution.  This procedure was followed in  \cite{Hartm2008}.  Following their calculations we find the equation for the free energy $f$ in the limit $N \rightarrow \infty$: 
\begin{eqnarray}
 f = f_{1} + f_{2},
\end{eqnarray}
with 
\begin{eqnarray}
\fl ik f_{1} &=  -\lim_{n \rightarrow 0} \frac{1}{ 2n} 
\sum_{\bsigma \btau} \Big[ -\Big( \frac{J_1k}{\sqrt{2}} \Big)^\alpha 
|\bsigma.\btau|^{\alpha}\left(1+i\gamma \rm{sign}(\bsigma\cdot\btau)\Phi\right) - i k J_0 ( \bsigma.\btau )  \Big]
P(\bsigma) P(\btau) \,\,, \label{eq:fEnergy1} \\
\fl ik f_{2} &=  \lim_{n \rightarrow 0} \frac{1}{ n}
\log{\biggl\{ \sum_{\bsigma} \exp\left[-\sum_{\btau} \Big[ -\Big( \frac{J_1k}{\sqrt{2}} \Big)^\alpha 
|\bsigma.\btau|^{\alpha}\left(1+i\gamma \rm{sign}(\bsigma\cdot\btau)\Phi\right) - i k J_0 ( \bsigma.\btau )  \Big]P(\btau) \right] \biggr\} }. \nonumber \\ \fl \label{eq:fEntro1}
\end{eqnarray}
The order parameter $P(\bsigma)$, with $\bsigma = \left(\sigma^1, \sigma^2, \ldots, \sigma^n\right)$, fulfills the self-consistent equation
\begin{eqnarray}
\fl P(\bsigma)&= \frac{\exp\left(\sum_{ \btau}P(\btau)\left[-ikJ_0\bsigma\cdot \btau -  \left|\frac{J_1k \bsigma\cdot\btau}{\sqrt{2}}\right|^{\alpha}\left(1+i\gamma \rm{sign}(\bsigma\cdot\btau)\Phi\right)\right]\right)}{\sum_{\bsigma}\exp\left(\sum_{ \btau}P(\btau)\left[-ikJ_0\bsigma\cdot \btau -  \left|\frac{J_1k \bsigma\cdot\btau}{\sqrt{2}}\right|^{\alpha}\left(1+i\gamma \rm{sign}(\bsigma\cdot\btau)\Phi\right)\right]\right)}.  
 \label{eq:PFinal}
\end{eqnarray}
We make 
the replica-symmetric (RS) ansatz, 
\begin{eqnarray}
 P(\bsigma) = \int dh W(h) \prod^n_{a=1}\frac{\exp\left(-ikh \sigma^a\right)}{2\cosh\left(-ik h \sigma^a\right)}, \label{eq:RS}
\end{eqnarray}
which defines the field distribution $W(h)$. Substitution of (\ref{eq:RS}) in (\ref{eq:PFinal}) gives:

\begin{eqnarray}
\fl
W(h) = \int \frac{ds}{2\pi} \exp{(i s h)} \exp\Bigg\{
- \int dh W(h) \int 
\frac{d\hat{J} dJ}{2 \pi}  \nonumber \\
\times
\left[ \left(\frac{J_1}{\sqrt{2}}\right)^{\alpha} |\hat{J}|^\alpha 
\Big(1 + i \gamma \Phi {\rm sign}(\hat{J}) \Big) + i J_0 \hat{J} \right] \exp\left[i \hat{J} J\right]
f(J,h,s) \Bigg\}. \label{defWh}
\end{eqnarray}
The function $f(J,h,s)$ is defined
as
\begin{equation}
f(J,h, s) \equiv \exp{\left( - \frac{i s}{\beta} {\rm atanh} \Big[ \tanh\left(\beta J\right) 
\tanh\left(\beta h\right)  
\Big]\right)} \label{funcdefin}.
\end{equation}
The analytic continuation of $T$ to real values
has been achieved by taking $k=i \beta$ 
at the end of the calculation. 
Using eqs.~(\ref{defI1})-(\ref{finalI2B}) 
from appendix B, we integrate  in eq.~(\ref{defWh}) over the $\hat{J}$ variable to obtain the following simplified expression for $W(h)$
\begin{eqnarray}
\fl
0 < \alpha &<& 1 : \nonumber \\ 
\fl
W(h)& =& \int \frac{ds}{2\pi} \exp{\Bigg{\{}i s h - i s J_0 m + 
\int dh W(h) \int_{-\infty}^{\infty} dJ  
\rho(J)
\Big[ f(J,h,s) -1   
\Big] \Bigg{\}}} , \label{eqWh1A}
\end{eqnarray}
\begin{eqnarray}
\fl
1 < \alpha < 2 : \nonumber \\ 
\fl
W(h) = \int \frac{ds}{2\pi} \exp{\Big( i s h - i s J_0 m \Big)} \nonumber \\
\times \exp{\Bigg{\{}
\int dh W(h) \int_{-\infty}^{\infty} dJ  
\rho(J)
\Big[ f(J,h, s) 
- f^{\prime}(0,h,s)  J  -1   \Big] \Bigg{\}}}, \label{eqWh1B}
\end{eqnarray}
where $f^{\prime}(0,h,s)= \frac{\partial f(J,h,s)}{\partial J}\Big{|}_{J=0}$. 
The distribution of couplings $\rho(J)$
is defined by eq.~(\ref{asympP}).
The RS magnetization $m$ and the RS 
spin-glass order-parameter $q$ are determined through
the averages
\begin{equation}
m = \int dh W(h) \tanh{(\beta h)}, \qquad 
q = \int dh W(h) \tanh^{2}{(\beta h)} . \label{eq:mQ}
\end{equation}

Only the large tail behavior of the distribution $P^{J_1, \gamma J_0}_\alpha$ appears in the equations~(\ref{eqWh1A}) and (\ref{eqWh1B}).
This could mean that the system exhibits a certain degree of universality: the thermodynamic behavior only depends on the large tail behavior of the coupling distribution $P(J)$. 
The distribution $\rho(J)$ is symmetric when $\gamma=0$, with
eqs.~(\ref{eqWh1A}) and (\ref{eqWh1B}) reducing to a
single equation, obtained previously in \cite{Hartm2008}. 

%%%%%%%%%%%%%%%%%%%%%%%%%%%%%%%%%%%%%%%%%%%%%%%%

\subsection{The normalization of the coupling distribution through a cutoff}

The main difficulty in eqs.~(\ref{eqWh1A}) and 
(\ref{eqWh1B}) concerns the normalization of $\rho(J)$ since
the integral $\int d J \rho(J)$
diverges for $\alpha < 2$.  Therefore, it is not possible to normalize the distribution.  This
invalidates the numerical calculation of $W(h)$ through  
the population dynamics algorithm \cite{Mez2001} because it is not possible to
sample random numbers from a non-normalizable distribution.

 In this subsection we propose
a simple procedure that allows to normalize 
$\rho(J)$ and to derive a self-consistent 
equation for $W(h)$ which is similar to the order-parameter equation of
FiC spin glasses on random 
graphs.  The numerical solution of this equation can be obtained through population
dynamics. 

The method consists of the insertion
of a temperature dependent cutoff $T \epsilon > 0$ in the integrals 
over $J$ occurring in eqs.~(\ref{eqWh1A}) and 
(\ref{eqWh1B}), splitting each of them into 
an integral around zero (from $-T \epsilon $ to $T \epsilon$) 
plus an integral over the couplings 
that satisfy $|J| > T \epsilon$. Assuming 
$T \epsilon \ll 1$, the 
integrals around zero can be analytically performed 
by expanding $f(J,h,s)$ around $J=0$ up to
order $\mathcal{O}(J^2)$, resulting in the following equations
\begin{eqnarray}
\fl
0 < \alpha < 1 : \nonumber \\ 
\fl
\int_{-\infty}^{\infty} dJ  \rho(J) \Big[ f(J,h,s) -1   \Big]=
- 2 i s \gamma  C_{\alpha}  \tanh{(\beta h)} 
\frac{(T \epsilon)^{1-\alpha}}{1-\alpha} 
- s^2 
C_{\alpha} \tanh^{2}{(\beta h)} \frac{(T \epsilon)^{2-\alpha}}{2-\alpha} \nonumber \\
+ \int_{-\infty}^{\infty} dJ \rho(J) 
\Big[ \Theta(J-T\epsilon) + \Theta(-J-T\epsilon) \Big]
\Big[ f(J,h,s) -1   \Big], \label{eqa}
\end{eqnarray}
\begin{eqnarray}
\fl
1 < \alpha < 2 : \nonumber \\ 
\fl
\int_{-\infty}^{\infty} dJ  \rho(J) \Big[ f(J,h,s)- f^{\prime}(0,h,s) J -1 \Big]= 
-  s^2 C_{\alpha} \tanh^{2}{(\beta h)} \frac{(T\epsilon)^{2-\alpha}}{2-\alpha} \nonumber \\
\fl \ \ \ \ \ \ \ + \int_{-\infty}^{\infty} dJ \rho(J) 
\Big[ \Theta(J-T\epsilon) + \Theta(-J-T\epsilon) \Big]
\Big[ f(J,h,s)-f^{\prime}(0,h,s) J -1   \Big].\label{eqb}
\end{eqnarray}
The symbol $\Theta(J)$ denotes the Heaviside step function: $\Theta(J)=1$ 
if $J > 0$ and $\Theta(J)=0$ otherwise.
We define the normalized distribution $P_\epsilon(J)$ in terms of $\rho(J)$
\begin{equation}
P_{\epsilon}(J) \equiv  \frac{\alpha (T\epsilon)^{\alpha}}{2 C_\alpha}  \rho(J) 
\Big[ \Theta(J-T\epsilon) + \Theta(-J-T\epsilon) \Big]  
. \label{normdistr}
\end{equation}
Subsituting eqs.~(\ref{eqa}) and (\ref{eqb}) in, respectively,
eqs.~(\ref{eqWh1A}) and (\ref{eqWh1B}) the integrals over $s$ can be
analytically calculated:
\begin{eqnarray}
 \fl W_{\epsilon}(h) = \sum^{\infty}_{k=0}\exp\left(-c\right)\frac{c^k}{k!} \int 
\left(\prod^k_{r=1}dh_rW_{\epsilon}(h_r)\right)\int \left(\prod^k_{r=1}dJ_r P_{\epsilon}(J_r)\right) 
\int \mathcal{D}z \nonumber \\ 
\fl \times \delta\left(h
- \tilde{J}_0m
-\beta^{-1}\sum^k_{r=1}
\atanh\Big[\tanh\left(\beta J_r\right)\tanh\left(\beta h_r\right) \Big]  
-\sqrt{2q\Delta} z\right), \label{eq:Distri}
\end{eqnarray}
where $\mathcal{D}z = (2 \pi)^{-\frac{1}{2}} \exp{(-z^2/2)}dz$ and:
\begin{eqnarray}
c= \frac{2 
C_{\alpha}}{\alpha (T\epsilon)^\alpha}, \qquad
\Delta =\frac{ 
(T\epsilon)^{2-\alpha} C_{\alpha}}{2-\alpha}, \label{eq:Delta}  \\ 
 \tilde{J}_0 = \left(J_0 
+ 2\gamma 
C_{\alpha}  \left[ \frac{(T \epsilon)^{1-\alpha}}{1-\alpha} \right] \right). \label{eq:JEffect}
\end{eqnarray}
To describe the thermodynamic behavior of L\'evy spin glasses one has to solve the set of equations (\ref{eq:mQ}) and (\ref{eq:Distri}) for $\epsilon\rightarrow 0$.

When we compare equation ~(\ref{eq:Distri}) with the order parameter equations of FiC systems \cite{MP87}, it describes
the effective field distribution 
of a FiC system of
Ising spins, in which the number of connections per site $k$ follows
a Poissonian distribution with connectivity $c$. The values
of the $k$ couplings attached to a site are drawn from the 
distribution $P_{\epsilon}(J)$, see eq. (\ref{normdistr}). In addition, the analytical 
calculation of the integrals over
the couplings that satisfy $|J| < T \epsilon$ yields an interaction with the global magnetization with effective strength $\tilde{J}_0$ and
an extra source of noise in eq.~(\ref{eq:Distri}), represented
by the Gaussian random variable $z$ with zero mean and variance 
$\Delta$.  The effective strength contains the shift parameter $J_0$ and a term linear in $\gamma$ corresponding with the center of the distribution of the couplings.  The interpretation of equation ~(\ref{eq:Distri}) is clear  : the effective field contains a Poissonian term coming from a finite number of strong bonds and a Gaussian term coming from an infinite number of weak bonds it interacts with.  
 One can perform the limit $\alpha\rightarrow 2$ to find the effective field $J_0 m + J_1\sqrt{q} z$, i.e. the RS solution of the SK model.  The equations (\ref{eq:mQ}) and (\ref{eq:Distri}) show explicitly how L\'evy spin glasses are a hybrid between FC and FiC models.  When one takes a Gaussian ansatz for the distribution $W_\epsilon$, equation (\ref{eq:Distri}) becomes in the limit $\epsilon\rightarrow 0$ equal to the result derived by Cizeau and Bouchaud \cite{Ciz1993}.

The population dynamics algorithm \cite{Mez2001} can be easily
adapted to solve numerically eq.~(\ref{eq:Distri}) and obtain $W_{\epsilon}(h)$.
The idea is to obtain numerical results for sufficiently small values
of $T \epsilon$ in a way that they can be extrapolated
for $T \epsilon \rightarrow 0$: the first two moments of the distribution already obtain their limiting values around $T\epsilon \lesssim 0.5$.  The equations become very hard to solve around $\alpha\approx 1.5$ because the mean connectivity $c$ has a maximum there.  For low values of $\alpha\lesssim 0.1$ population dynamics becomes inaccurate because of numerical imprecisions due to the larger tails of the coupling distribution.

\begin{figure}[h!]
\begin{center}
\includegraphics[angle=-90, width=.6 \textwidth]{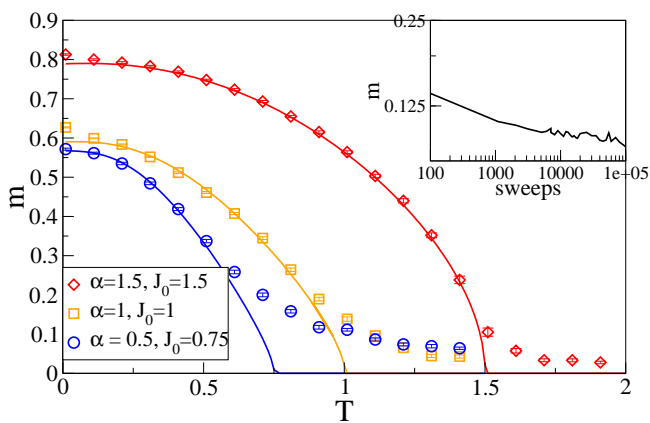}
\caption{The magnetization $m$ as a function of the temperature $T$ for several values of $\alpha$ and $J_0$. Simulation results (markers) are 
compared with results from the theory (solid lines) for $J_1=1$, $\gamma = 0$.  At low temperatures 
theory and simulations are in good agreement. Because of the increase in the equilibration 
time around the critical temperature the results from simulation overestimate the magnetization.  The inset confirms this: it shows the value of the magnetization as a function of the number of Monte Carlo sweeps for $\alpha=0.5, J_0 = 0.75$ and $T = 1$.}\label{fig:simCompareLast}
\end{center}
\end{figure}

In figure \ref{fig:simCompareLast} we compare the solution of equations (\ref{eq:mQ}) and (\ref{eq:Distri}) with Monte-Carlo simulations.  We simulated a L\'evy spin glass using the algorithm described in \cite{Hartm2008} without the parallel tempering.  The algorithm contains two update rules: single spin flip updates as usually done in Metropolis algorithms and updates of clusters of spins connected through strong bonds.  For low temperatures we find a good agreement between the simulations and the theory.  Around the critical temperature the magnetization obtained by the simulations is larger than the one predicted by the theory.  The reason for this difference is that the simulations equilibrate very slowly.   Indeed, as shown in the inset of figure \ref{fig:simCompareLast} the magnetization decays as a power law as a function of the number of Monte-Carlo sweeps. The presence of strong bonds slows down the dynamics since the effect becomes larger for smaller values of $\alpha$.  For very low temperatures the simulation results for the magnetization deviate from those of the RS result.  The RS ansatz (\ref{eq:RS}) is invalid for very low temperatures, see section \ref{sec:phase}.  

\begin{figure}[h!]
\begin{center}
\includegraphics[angle=-90, width=.6 \textwidth]{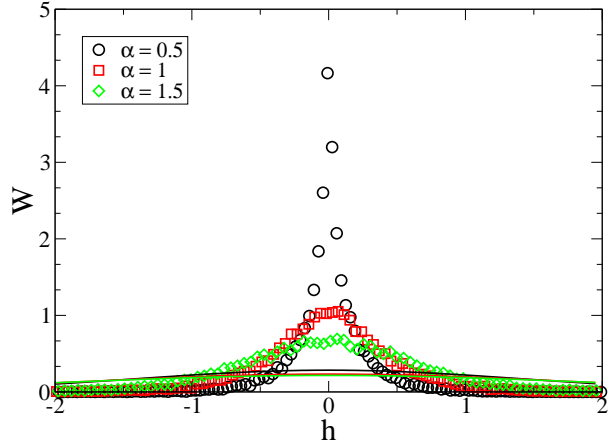}
\caption{The distribution of effective fields with $J_0= \gamma = 0$, $J_1=1$, $T = 0.4$ and several values of $\alpha$.  
The markers give the distributions according to equation (\ref{eq:Distri}) while the solid lines are obtained through the theory of \cite{Ciz1993}.  All the moments of the distributions are finite.  Therefore we have leptokurtic distributions which are neither Gaussian nor L\'evy distributions.}\label{fig:distri}
\end{center}
\end{figure}

In figure \ref{fig:distri} we plotted the solution to the self-consistent equation (\ref{eq:Distri}) for different values of $\alpha$.  The result is compared with the Gaussian ansatz (solid lines), used in \cite{Ciz1993}.  The difference between both approaches is clear.  For $\alpha\rightarrow 2$ the distribution of fields becomes more and more Gaussian.  For $\alpha<2$ the distribution of fields is not L\'evy but leptokurtic distributions where the moments converge to a finite value as a function of the size of the population.  Leptokurtic distributions have a smaller kurtosis than a Gaussian distribution with the same variance.  

%%%%%%%%%%%%%%%%%%%%%%%%%%%%%%%%%%%%%%%%%%%%%%%%%%%%%%%%%%%%%%%%%

\section{Cavity method}\label{sec:cavity}
We derive the self-consistent eqs.~(\ref{eq:Distri}) for $W(h)$ through the cavity method.  In contrast with \cite{Ciz1993} we only apply the CLT to the field coming from the weak bonds, i.e. bonds smaller than the cutoff $T\epsilon$.  The bonds larger than $T\epsilon$ form a backbone graph of strong bonds which is treated as a FiC system.  For $\epsilon\rightarrow \infty$ we find back the results of \cite{Ciz1993}. For $\epsilon\rightarrow 0$ we expect to find the RS behavior of the spin glass.

The marginal $P_i(\sigma_i) \equiv \sum_{\left\{\sigma_j\right\}_{j=1..N}\setminus \sigma_i} P\left(\left\{\sigma_j\right\}_{j=1..N}\right)$ of the Gibbs distribution $P\left(\left\{\sigma_j\right\}_{j=1..N}\right) \sim \exp\left[-\beta H\left(\left\{\sigma_j\right\}_{j=1..N}\right)\right]$ can be written as
\begin{eqnarray}
 P_i(\sigma_i) &\sim \sum_{\left\{\sigma_j\right\}_{j=1..N}\setminus \sigma_i}P^{(i)}\left(\left\{\sigma_j\right\}_{j=1..N}\setminus \sigma_i\right)\exp\left(\sum_{k}J_{ki}\sigma_j\sigma_i\right) ,
\end{eqnarray}
with $P^{(i)}\left(\left\{\sigma_j\right\}_{j=1..N}\setminus \sigma_i\right)$ the Gibbs distribution on the cavity graph $G^{(i)}$.  The cavity graph is the subgraph of the original graph $G$ where one removed the $i$-th spin and all of its interactions with the other spins.  
We assume that the probability distribution on the cavity graph factorizes \cite{par1987}: 
\begin{eqnarray}
P^{(i)}\left(\left\{\sigma_j\right\}_{j=1..N}\setminus \sigma_i\right) = \prod_{j(\neq i)} P^{(i)}_j(\sigma_j) .
\end{eqnarray}
This factorization is valid when there is one pure phase in the system.
The set $\overline{\omega}^{(i)}$ of all weak bonds and the set  $\omega^{(i)}$ of all strong bonds are defined through: 
\begin{eqnarray}
\overline{\omega}^{(i)}  \equiv \left\{j\in \mathbb{N}\cap [1,N]|J_{ij}< T \epsilon\right\},\\
\omega^{(i)} \equiv (\mathbb{N}\cap [1,N]) \setminus \overline{\omega}^{(i)}.
\end{eqnarray}
The cavity fields $h^{(i)}_j$ and $g^{(i)}_j$ are defined through: 
\begin{eqnarray}
h^{(i)}_j \equiv \sum_{\sigma}\frac{\sigma}{2}\log\left(P^{(i)}_j(\sigma)\right)&\rm{if}& j\in \omega^{(i)}, \\ 
g^{(i)}_j \equiv \sum_{\sigma}\frac{\sigma}{2}\log\left(P^{(i)}_j(\sigma)\right) &\rm{if}& j \in \overline{\omega}^{(i)}.
\end{eqnarray}
The marginal $P^{(j)}_i$ of the $i$-th spin on the cavity graph $G^{(j)}$ is equal to:   
\begin{eqnarray}
 \fl  P^{(j)}_i(\sigma_i) \sim \prod_{k\in \overline{\omega}^{(i)}\setminus j}\sum_{\tau}\exp\left(\beta J_{ki}\sigma_i \tau + \beta g^{(i)}_k \tau\right) \prod_{k\in \omega^{(i)} \setminus j}\sum_{\tau}\exp\left(\beta J_{ki}\sigma_i \tau + \beta h^{(i)}_k \tau\right). \label{eq:Pj}
\end{eqnarray}
We used the notation $h^{(j)}_i$ for cavity fields where one removed a site $j$ connected with $i$ through a strong bond and the notation $g^{(j)}_i$ for fields where the site $j$ was connected with $i$ through a weak bond .
We thus find the following set of closed equations in the cavity fields $h^{(j)}_i$ and $g^{(j)}_i$: 
\begin{eqnarray}
\fl  g^{(j)}_i = z^{(j)}_i+ \beta^{-1}\sum_{k\in\omega_i}\atanh\left(\tanh\left(\beta h^{(i)}_k\right) \tanh\left(\beta J_{ki}\right)\right), \label{eq:g}\\ 
\fl  h^{(j)}_i = z_i + \beta^{-1}\sum_{k\in\omega_i\setminus j}\atanh\left(\tanh\left(\beta h^{(i)}_k\right) \tanh\left(\beta J_{ki}\right)\right), \label{eq:h}
\end{eqnarray}
where we defined a third field containing the contributions from the weak bonds
\begin{eqnarray}
 z^{(j)}_i =  \beta^{-1}\sum_{k\in \overline{\omega}_i\setminus j }\atanh\left(\tanh\left(\beta g^{(i)}_k\right) \tanh\left(\beta J_{ki}\right)\right). \label{eq:z}
\end{eqnarray}
In the limit $N\rightarrow \infty$ we can remove the $j$ dependency in the fields $z^{(j)}_i$ and $g^{(j)}_i$ because the sum over the weak bonds ($k\in \overline{\omega_i}$) contains an infinite number of terms.

To take the disorder average over the couplings we define the following distributions:  
\begin{eqnarray}
 W_{\rm{w}}(g) \equiv \overline{\frac{\sum^{N}_{i=1}\delta\left(g-g_i\right)}{N}}, \label{eq:Pwg}\\ 
 W_{\rm{s}}(h) \equiv \overline{\frac{\sum^{N}_{i=1}\sum_{j\in \omega_i } \delta\left(h-h^{(j)}_i\right)}{\sum^{N}_{i=1}\sum_{j\in \omega_i }}},\label{eq:Psh} \\ 
W_{\rm g}(z) \equiv \overline{\frac{\sum^{N}_{i=1}\delta(z-z_i)}{N}}.
\end{eqnarray}
 We treat the $z$-fields as a sum of infinitely many random variables on which we can apply the CLT:    
\begin{eqnarray}
 W_{\rm g}(z) = \frac{1}{\sqrt{4\pi \Delta q}}\exp\left(-\frac{(z-\tilde{J_0} m)^2}{4\Delta q}\right),
\end{eqnarray}
with $\Delta$ and $\tilde{J}_0$ as defined in eqs.~(\ref{eq:Delta}) and (\ref{eq:JEffect}).  The parameters $m$ and $q$ determine, respectively, the mean and the variance of the Gaussian distribution $W_{\rm g}(z)$. 
Here is the important difference with \cite{Ciz1993} where the CLT is applied on all bonds, also on the strong ones.  

From  eq. (\ref{eq:z}) one finds, for $N\rightarrow \infty$ and $\epsilon \ll 1$ the following expressions for the mean $m$ and the variance $q$ 
\begin{eqnarray}
  m = \left(\tilde{J}_0\right)^{-1}\left(N\int^{T\epsilon}_{-T\epsilon} dJ P^{J_1, \gamma, J_0 }_{\alpha}(J)  J\right) \int dg \tanh(\beta g)W_{\rm w}(g) ,  \label{eq:coupl1}\\  
q =  \left(2\Delta\right)^{-1}\left(N\int^{T\epsilon}_{-T\epsilon} dJ P^{J_1, \gamma, J_0 }_{\alpha}(J)  J^2\right) \int dg\tanh^2(\beta g) W_{\rm w}(g). \label{eq:coupl2}
\end{eqnarray}
The integrals over the couplings in eqs.~(\ref{eq:coupl1}) and (\ref{eq:coupl2}) can be calculated using analogous methods as used to derive the integrals in appendix B:  
\begin{eqnarray}
N \int^{T\epsilon}_{-T\epsilon} dJ P^{J_1, \gamma, J_0 }_{\alpha}(J)  J = \tilde{J}_0, \\
N\int^{T\epsilon}_{-T\epsilon} dJ P^{J_1, \gamma, J_0 }_{\alpha}(J)  J^2 = 2\Delta .
\end{eqnarray}
Using the definitions of the distributions $W_{\rm{w}}(g)$  and  $W_{\rm{s}}(h)$ in eqs.~(\ref{eq:Pwg}) and (\ref{eq:Psh}) we get
\begin{eqnarray}
\fl W_{\rm s}(h) &=  \sum^{\infty}_{k=0}\frac{p_{\rm poiss}(k;c)k}{c} \prod^{k-1}_{r=1} \int dh_r W_{\rm s}(h_r)\int\prod^{k-1}_{r=1}dJ_r P_{\epsilon}(J_r)\int dz W_{\rm g}(z)
\nonumber \\
\fl& \delta\left(h - z - \beta^{-1}\sum^{k-1}_{r=1}\atanh\left(\tanh(\beta h_r)\tanh(\beta J_r )\right)\right) , \label{eq:PsSelfc}\\ 
\fl W_{\rm w}(g) &=  \sum^{\infty}_{k=0}p_{\rm poiss}(k;c)\int  \prod^{k}_{r=1} dh_r W_{\rm s}(h_r)\int\prod^{k}_{r=1}dJ_rP_{\epsilon}(J_r)\int dz W_{\rm g}(z)
\nonumber \\
\fl& \delta\left(g - z - \beta^{-1}\sum^{k}_{r=1}\atanh\left(\tanh(\beta h_r)\tanh(\beta J_r )\right)\right). \label{eq:PwSelfc}
\end{eqnarray}
The mean connectivity $c$ is given by
\begin{eqnarray}
 c = \lim_{N\rightarrow\infty}\frac{\sum^N_{i=1}|\omega^{(i)}(\epsilon)|}{N} =  \int^{\infty}_{T\epsilon} dJ \rho(J) +  \int^{-T\epsilon}_{-\infty} dJ \rho(J)  =\frac{2C_\alpha}{\alpha(T\epsilon)^{\alpha}},
\end{eqnarray}
with $\rho(J)$ the large tail behavior as defined in (\ref{asympP}).  We use the following property of Poissonian distributions: $\frac{1}{c}p_{\rm poiss}(k; c)k = p_{\rm poiss}(k-1; c)$ to find  $W_{\rm w}(g) =  W_{\rm s}(g) = W_{\epsilon}(g)$, i.e. the solutions to (\ref{eq:PsSelfc}) and (\ref{eq:PwSelfc}) are the same as the solution $W_{\epsilon}$ of (\ref{eq:Distri}).  Indeed, eqs.~(\ref{eq:PwSelfc}) combined with (\ref{eq:coupl1}) and (\ref{eq:coupl2}) are identical to eqs.~(\ref{eq:mQ}) and (\ref{eq:Distri}) derived with the replica method.   From the cavity approach the importance of the CLT in L\'evy spin glasses becomes clear: the couplings have a divergent variance, therefore one can not apply the CLT as done in \cite{Ciz1993}.  We remark that the effective coupling $\tilde{J}_0$ and the parameter $2\Delta$ appearing in the replica method are the mean and the variance of the weak couplings.  The distribution $W_{\epsilon}(h)$ in equations (\ref{eq:Distri}) is the distribution of the cavity fields propagating along the backbone graph of strong bonds.

%%%%%%%%%%%%%%%%%%%%%%%%%%%%%%%%%%%%%%%%%%%%%%%%%%%%%%%%%%%%%%%%%
\section{Stability of the replica symmetric ansatz}\label{sec:RSB}
%%%%%%%%%%%%%%%%%%%%%%%%%%%%%%%%%%%%%%%%%%%%%%%%%%%%
As is known \cite{TH1978}, the RS ansatz introduced in (\ref{eq:RS}) is unstable at low temperatures.  It is possible to calculate the regions of stability by using the two replica method, first introduced for FiC models in \cite{Kwon1991}.  This method allows us to study local and non-local replica symmetry breaking (RSB) effects.  For models on graphs a relevant instability condition is proved rigorously in \cite{aldous2005}.   It determines the region where the message passing algorithms stop to converge, see for example the discussion in \cite{Neri}.   

We start by considering two uncoupled replicas.  Both replicas fulfill the equations (\ref{eq:Pj})-(\ref{eq:z}).   The replicas only get coupled when we take the average over the graph instance.  Indeed, the effective field distribution of two sets of uncoupled spins on the same graph with the same couplings is given by:  
\begin{eqnarray}
\fl W_{\epsilon}(h^1, h^2) &=  \sum^{\infty}_{k=0}\frac{p_{\rm poiss}(k;c)k}{c} \prod^{k-1}_{r=1} \int dh^1_rdh^2_r W_{\epsilon}(h^1_r, h^2_r)\int\prod^{k-1}_{r=1}dJ_r P_{\epsilon}(J_r)\int dz^1dz^2 W_{\rm g}(z^1, z^2)
\nonumber \\
\fl& \times \delta\left(h^1 - z^1 - \beta^{-1}\sum^{k-1}_{r=1}\atanh\left(\tanh(\beta h^1_r)\tanh(\beta J_r )\right)\right)
\nonumber \\
\fl & \times \delta\left(h^2 - z^2 - \beta^{-1}\sum^{k-1}_{r=1}\atanh\left(\tanh(\beta h^2_r)\tanh(\beta J_r )\right)\right)
. \label{eq:PsSelfc22} 
\end{eqnarray}
We assume again that we can apply the CLT on the $z$-fields: 
\begin{eqnarray}
 \fl W_{\rm g}(z^1, z^2) 
\nonumber \\
\fl 
= \frac{1}{4\Delta \pi \sqrt{q^1q^2 (1-\rho^2)}}\exp\left(-\frac{1}{2(1-\rho^2)}\left(\frac{(z^1-\tilde{J}_0m^1)^2}{2\Delta q^1} + \frac{(z^2-\tilde{J}_0m^2)^2}{2\Delta q^2}\right)\right)
\nonumber \\
\fl \times\exp\left(
\frac{\rho (z^1-\tilde{J}_0m^1)(z^2-\tilde{J}_0m^2)}{2(1-\rho^2)\Delta \sqrt{q^1q^2}}\right).
\end{eqnarray}

The order parameters become
\begin{eqnarray}
  m^1 =  \int dg^1dg^2 \tanh(\beta g^1)W_{\epsilon}(g^1, g^2) , \\  
 q^1 =  \int dg^1dg^2\tanh^2(\beta g^1) W_{\epsilon}(g^1, g^2), \\ 
 m^2 =  \int dg^1dg^2 \tanh(\beta g^2)W_{\epsilon}(g^1, g^2) , \\  
q^2 =  \int dg^1dg^2\tanh^2(\beta g^2) W_{\epsilon}(g^1, g^2), \\
\rho \sqrt{q^1q^2} = \int dg^1dg^2\tanh(\beta g^2)\tanh(\beta g^1) W_{\epsilon}(g^1, g^2).
\end{eqnarray}
In the limit $\alpha\rightarrow 2$ we find $W_{\epsilon}(h^1, h^2) = W_{\rm g}(h^1, h^2)$.  An expansion around the RS solution $m^1=m^2=m$, $q^1=q^2=q$ and $1-|\rho|\sim\mathcal{O}(\delta)$, with $\delta \ll1$, leads to the following stability condition: 
\begin{eqnarray}
  \fl \beta^{-2} = \int^{+\infty}_{-\infty} \frac{du}{\sqrt{2\pi}}\exp\left(- \frac{u^2}{2}\right)\sech^4\left(\beta \sqrt{q} u+ \beta J_0m\right).\label{eq:AT}
\end{eqnarray}
The parameters $(q, m)$ in (\ref{eq:AT}) are, respectively, the overlap parameter and the magnetization of the SK model. 
Equation (\ref{eq:AT}) is precisely the AT line of the SK model, see \cite{TH1978}.

\section{Phase Diagram}\label{sec:phase}
The system shows three phases which depend on the values of the order parameters $m$ and $q$ defined in (\ref{eq:mQ}): a paramagnetic phase (P) with $m=q=0$, a ferromagnetic phase (F) with $m>0, q>0$ and a spin-glass phase (SG) with $m=0, q>0$. \begin{figure}[h!]
\begin{center}
\includegraphics[angle=-90, width=.6 \textwidth]{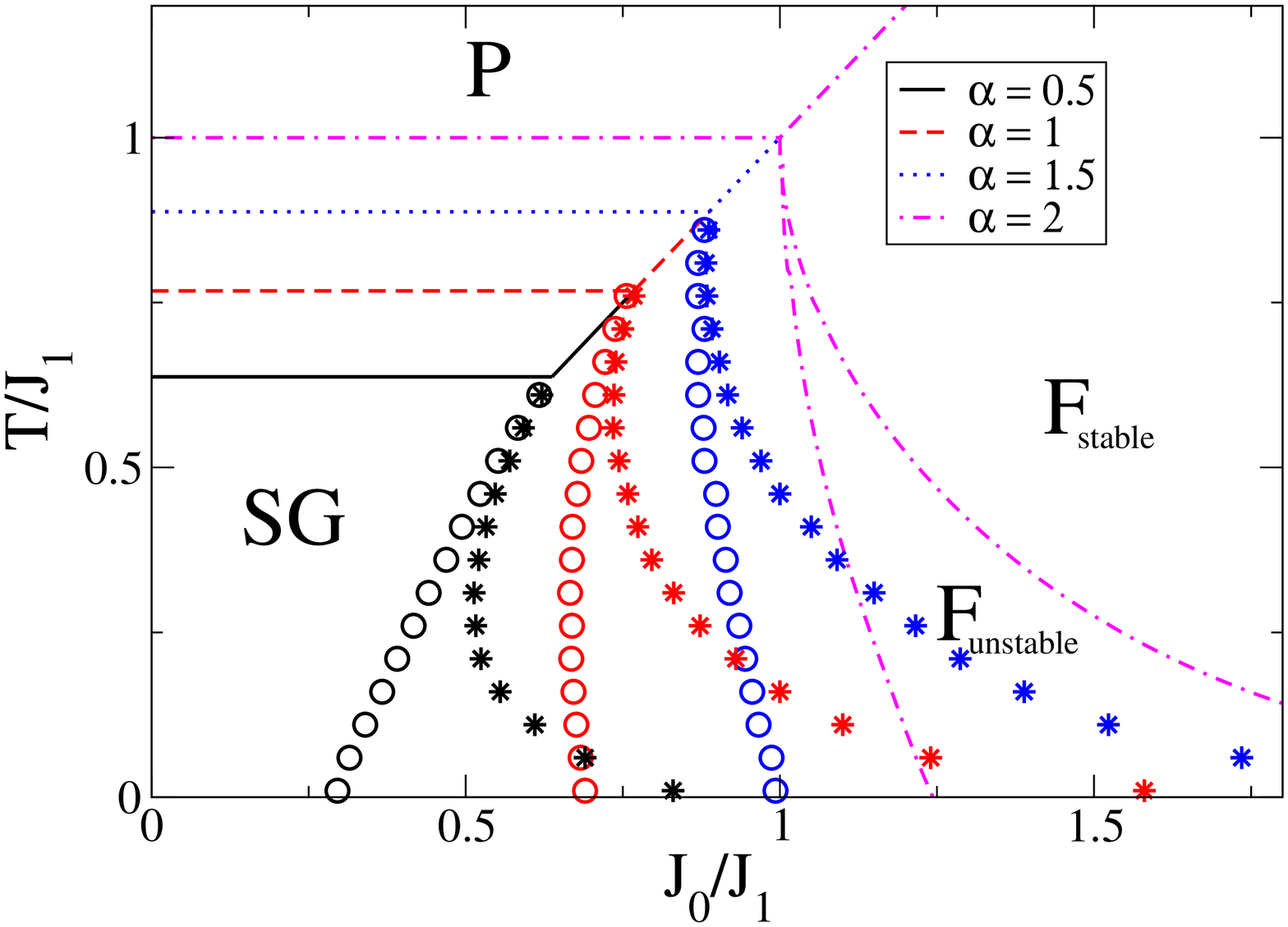}
\caption{The $(T/J_1, J_0/J_1)$ phase diagram for several values of $\alpha$ and with a skewness $\gamma =0$. Three phases appear: P (paramagnetic), F (ferromagnetic) and SG (spin glass). The circles present the SG-F transitions and the stars indicate where the F phase becomes stable against replica symmetry breaking.  For $\alpha=2$ the phase diagram coincides with that of the SK model.}\label{fig:phaseDiagram}
\end{center}

\begin{minipage}{.45\textwidth}
\begin{center}
\includegraphics[angle=-90, width = 1\textwidth]{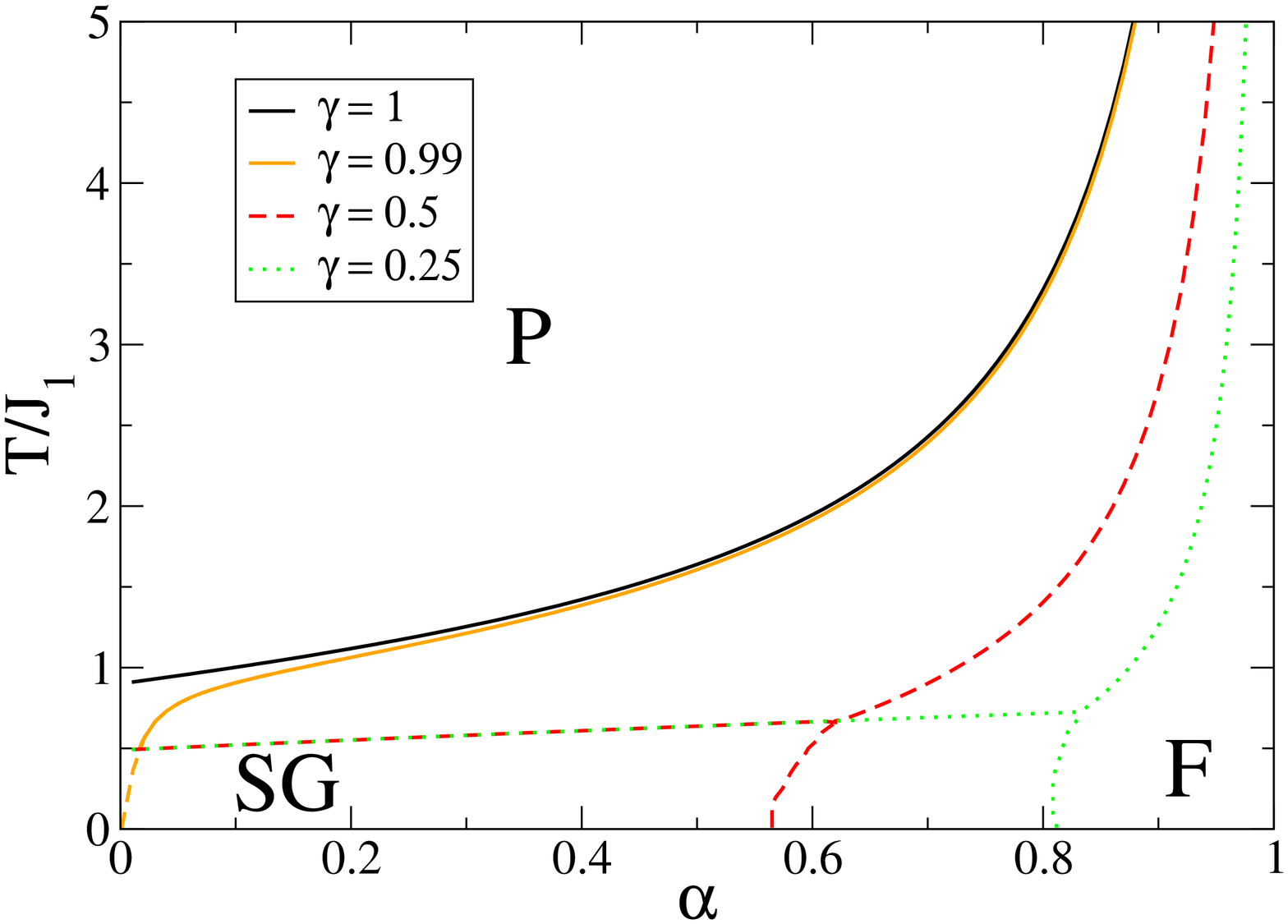}
\end{center}
\end{minipage}
\hfill
\begin{minipage}{.45\textwidth}
\begin{center}
\includegraphics[angle=-90,width= 1\textwidth]{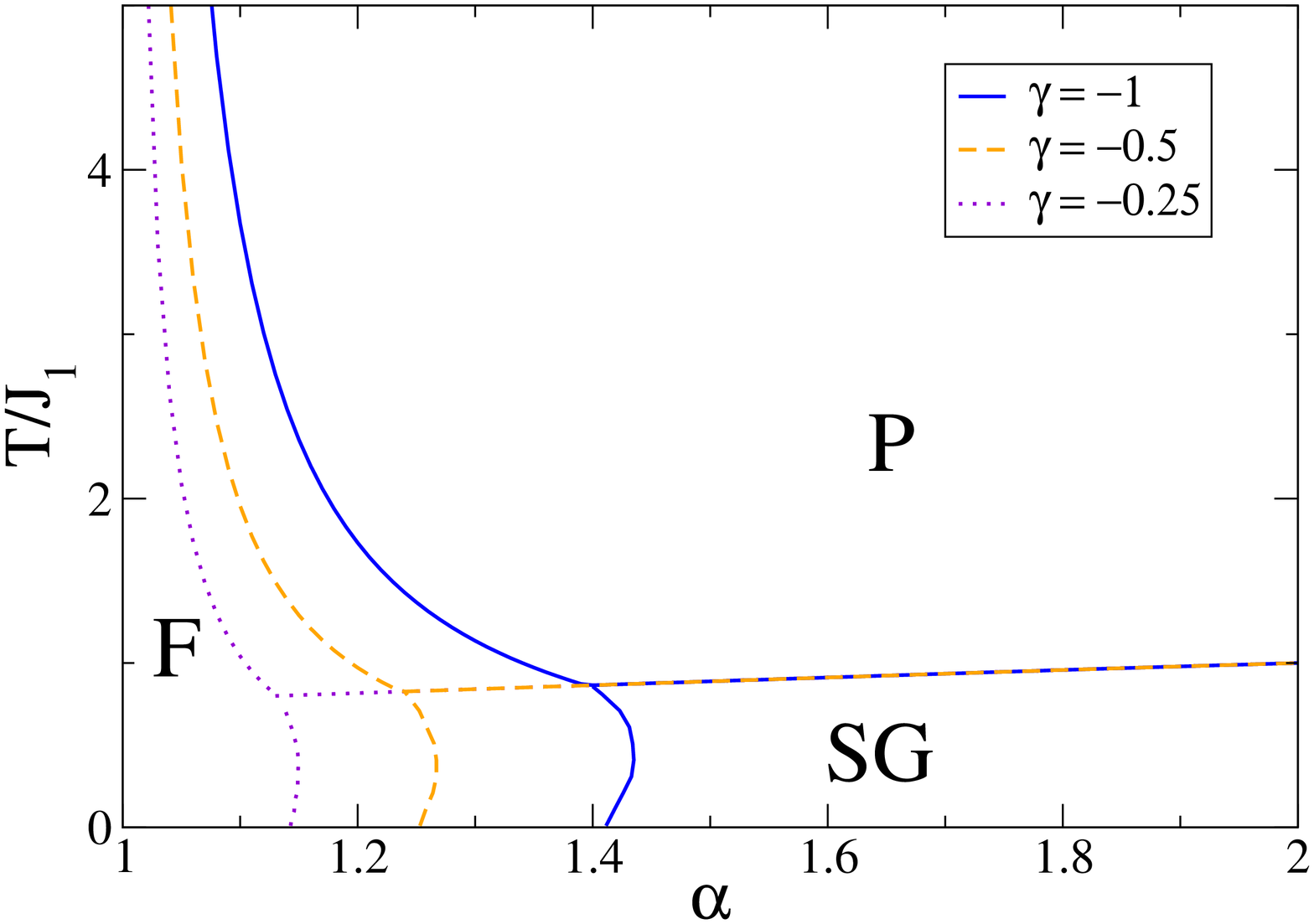}
\end{center}
\end{minipage}
\caption{The $(T/J_1, \alpha)$ phase diagrams for different values of the skewness $\gamma$ and a shift $J_0=0$.  The figure on the left gives the phase diagram for $\gamma>0$ while the figure on the right gives the phase diagram for $\gamma<0$.  For $\gamma=1$ and $\alpha<1$ there is no SG phase.  The $P-SG$ transition is independent of $\gamma$.  For $\gamma=0.99$ the dashed part of the transition line does not present the SG-F transition but the instability line of the P phase with respect to the F phase.
%For $\gamma=0.99$ the dashed line is a guide for the eye for the SG to F phase transition.
}
\label{fig:phaseDiagramGamma}
\end{figure} The ferromagnetic phase contains a region stable to RSB effects ($F_{\rm stable}$) and a region unstable to RSB effects $(F_{\rm unstable})$. 

 The P-F and P-SG transitions are determined using an expansion of the self-consistent eq. (\ref{eq:Distri}) around the paramagnetic solution $W_{\epsilon}(h) =\delta(h)$.  For $\gamma=0$ we find the same bifurcation lines as derived in \cite{Hartm2008}.  To determine the SG-F transition and the $F_{\rm unstable}$ to $F_{\rm stable}$ transition one has to solve numerically, respectively, eqs. (\ref{eq:Distri}) and   (\ref{eq:PsSelfc22}) with for instance a population dynamics algorithm.

In figure \ref{fig:phaseDiagram} the different phases in the $(J_0/J_1,T/J_1)$ phase diagram are presented for a skewness $\gamma=0$ and several values of $\alpha$.  The open circles present the SG-F transitions and the stars mark the points where the F phase becomes stable with respect to RSB.  These results generalize the phase diagram obtained in the seminal paper of Sherrington and Kirkpatrick \cite{SH1975} to coupling distributions with a large tail.   For $\gamma=0$ the P-F transition is independent of $\alpha$.   When $\alpha$ increases the SG phase increases in favor of a  smaller F phase.  The RSB effects decrease when $\alpha$ decreases: indeed the $F_{\rm unstable}$ becomes smaller and the reentrance effect in the SG-F phase transition line diminishes and finally disappears.  This is related to the decrease of frustration due to the presence of stronger bonds that dominate the systems behavior.  We did not find a replica symmetric SG phase (i.e., a SG phase stable with respect to RSB), contrary to the conjecture made in \cite{Ciz1993}.  Replica symmetry breaks continuously at the SG transition similar to the behavior of the SK model.  We did not find any further evidence for the conjecture in \cite{Ciz1993} that replica symmetry restores at $T=0$.  

In figure \ref{fig:phaseDiagramGamma} we present the $(T/J_1, \alpha)$-phase diagram for different values of $\gamma$ and $J_0=0$.  
We consider the following regions: 
\begin{itemize}
\item $\gamma>0$ and $\alpha<1$ (left figure): the F phase increases and the SG phase decreases as a function of increasing $\gamma$.  The SG phase disappears at $\gamma=1$.  For values of $\gamma$ very close to one the SG phase is only present for very small values of $\alpha$. The transition temperature between the P and F phase becomes infinite for $\alpha\rightarrow 1^-$.
\item $\gamma<0$ and $\alpha>1$ (right figure): the F phase decreases and the SG phase increases as a function of increasing $\gamma$.  The transition temperature between the P and F phase becomes infinite for $\alpha\rightarrow 1^+$.
\item $\gamma>0$ and $\alpha>1$ (not shown): there is no F phase but a P and SG phase. 
\item $\gamma<0$ and $\alpha<1$ (not shown): there is no F phase but a P and SG phase.
\end{itemize}
We have some additional remarks: The transition temperature becomes very large for $\alpha\rightarrow 1^{\pm}$ (for, respectively, $\gamma<0$ and $\gamma>0$) because the effective coupling $\tilde{J}_0\rightarrow \infty$.   There is no SG phase for $\gamma=1$ and $\alpha<1$ because there are no negative couplings, only the P-F transition occurs.  The P-SG transitions coincide for different values of $\gamma$.  For low values of $\alpha$ the results of population dynamics become inaccurate because of numerical imprecisions when dealing with a broad range of coupling values.  In this case we used the instability line of the P phase with respect to the F phase as  of the location of the SG-F transition.

\section{Entropy}\label{sec:entropy}
It is possible to calculate the free energy from the saddle point equations.  We use the RS ansatz and we introduce again a cutoff $\epsilon$.  The entropy is given by $s= \beta^2\frac{\partial f}{\partial \beta} = \lim_{\epsilon\rightarrow 0}s(\epsilon)$ with $s(\epsilon)$:
\begin{eqnarray}
 s(\epsilon) =   \beta^2 \frac{\Delta}{2}\left(1-q^2\right) - \beta^2 \Delta \left(1 - q\right)+ s_{\rm site}(\epsilon)-\frac{c}{2}s_{\rm link}(\epsilon).  \label{eq:entropy}
\end{eqnarray}
The quantity $s_{\rm link}$ is equal to
\begin{eqnarray}
\fl s_{\rm link}(\epsilon) &=  -\int dh dh^{\prime} W_{\epsilon}(h) W_{\epsilon}(h^{\prime})
\int dJ P_{\epsilon}(J) 
\nonumber \\ 
\fl &
\sum_{\sigma, \tau}\frac{  \exp\left(\beta J\sigma\tau + \beta h \sigma + \beta h'\tau\right)}{\sum_{\sigma, \tau}\exp\left(\beta J\sigma\tau + \beta h \sigma + \beta h'\tau\right)}  \log\left(\frac{  \exp\left(\beta J\sigma\tau + \beta h \sigma + \beta h'\tau\right)}{\sum_{\sigma, \tau}\exp\left(\beta J\sigma\tau + \beta h \sigma + \beta h'\tau\right)}  \right),
\end{eqnarray}
and $s_{\rm site}$ reads
\begin{eqnarray}
\fl s_{\rm site}(\epsilon)& = -\sum_{k=0}^{\infty} \frac{c^k e^{-c}}{k!}
\prod_{l=1}^{k} \biggl[\int dh_l W_{\epsilon}(h_l) 
\int dJ_l P_{\epsilon}(J_l)  \biggr] \int \mathcal{D}z 
\nonumber \\
\fl &\sum_{\sigma; (\tau_1, \tau_2, \cdots, \tau_k)}\frac{  \exp\left((\beta J_0 m +\sqrt{2q\Delta}z)\sigma \right) \prod^k_{\ell=1}\left(\exp\left(\beta J_\ell\tau_\ell\sigma\right) \exp\left(\beta h_\ell \tau_\ell\right)\right)}{\sum_{\sigma; (\tau_1, \tau_2, \cdots, \tau_k)} \exp\left((\beta J_0 m +\sqrt{2q\Delta}z)\sigma \right) \prod^k_{\ell=1}\left(\exp\left(\beta J_\ell\tau_\ell\sigma\right) \exp\left(\beta h_\ell \tau_\ell\right)\right)}
\nonumber \\ 
\fl &\log\left[\frac{ \exp\left((\beta J_0 m +\sqrt{2q\Delta}z)\sigma \right) \prod^k_{\ell=1}\left(\exp\left(\beta J_\ell\tau_\ell\sigma\right) \exp\left(\beta h_\ell \tau_\ell\right)\right)}{\sum_{\sigma; (\tau_1, \tau_2, \cdots, \tau_k)} \exp\left((\beta J_0 m +\sqrt{2q\Delta}z)\sigma \right) \prod^k_{\ell=1}\left(\exp\left(\beta J_\ell\tau_\ell\sigma\right) \exp\left(\beta h_\ell \tau_\ell\right)\right)}\right] .
\end{eqnarray}
For $\alpha\rightarrow 2$ one gets precisely the entropy of the SK model \cite{SK1978}.
The entropies $s_{\rm site}$ and $s_{\rm link}$ correspond with the entropy differences when performing, respectively, a site addition and a link addition on the backbone graph of strong bonds, see \cite{Mez2003, Mez2001}.  Similar to the form of the self-consistent equation (\ref{eq:Distri}) for $W_{\epsilon}(h)$ we find that the entropy as given by equation (\ref{eq:entropy}) corresponds to the entropy of an Ising model on a Poissonian graph with mean connectivity $c$, a distribution of the bonds $P_\epsilon$ and an extra Gaussian noise $z$.

\begin{figure}[h!]
\begin{center}
\includegraphics[angle=-90, width=.6 \textwidth]{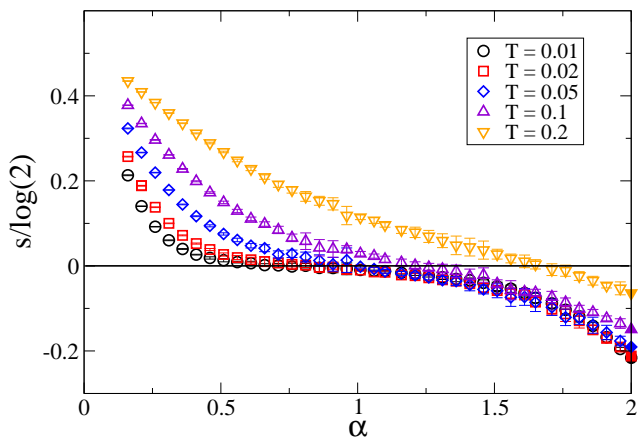}
\caption{The entropy $s$ as a function of the exponent $\alpha$ for different values of the temperature $T$, $J_0=0$, $\gamma=0$ and $J_1=1$.  The filled markers at $\alpha=2$ show the SK values.  The entropy converges to the SK value for $\alpha\rightarrow 2$.}\label{fig:entropy}
\end{center}
\end{figure}

We plotted the entropy $s$ as a function of $\alpha$ in figure \ref{fig:entropy}.  From this figure we see that the entropy gets less negative, for $T\rightarrow 0$.  We find that for smaller values of $\alpha\lesssim1$ the entropy becomes eventually zero for $T\rightarrow 0$.  This is consistent with a decrease of RSB effects when $\alpha$ decreases.

\section{Conclusion}\label{sec:conclusion}
In this paper we have shown how to derive the phase diagrams of L\'evy spin glasses where the couplings between the spins are drawn from a distribution with power law tails characterized by an exponent $\alpha$.  These models are known to have a finite number of strong bonds of order $\mathcal{O}(1)$ and an infinite amount of weak bonds of order $\mathcal{O}(N^{-1/\alpha})$.   The crucial difference with previous works \cite{Ciz1993} and \cite{Hartm2008} is that we derive the phase diagrams, the entropy and the stability against replica symmetry of L\'evy spin glasses without using the Gaussian assumption for the distribution of fields.  We have neither found evidence for a replica symmetric spin-glass phase, nor for a restoration of the replica symmetry at zero temperature, contrary to the conjecture in \cite{Ciz1993}.  We have solved the problem using the replica and the cavity method within, respectively, the replica symmetric assumption and the assumption of one pure phase.  The resultant effective equations for the distribution of cavity fields show clearly the hybrid character of the model being a mixture between a finite connectivity model and a fully connected model.    

The phase transitions are qualitatively similar to the ones found in the SK model. 
Large tails do influence quantitatively the phase diagram: the L\'evy spin-glass model becomes more stable with respect to replica symmetry breaking and the SG phase decreases when the tails get larger.  Moreover, the reentrance effects in the replica symmetric phase diagram disappear for $\alpha\lesssim 1$.  The replica symmetry breaking transitions are all continuous.  The skewness $\gamma$ in the L\'evy distribution can have a big influence on the size of the F phase.  For $\alpha\rightarrow 2$ the effective distribution of fields becomes Gaussian and we have found back the results of the SK model.  For $\alpha<2$ this distribution is neither L\'evy nor Gaussian, but a distribution with finite moments and a kurtosis smaller than a Gaussian with the same variance. 
  
\ack
One of the authors (F. L. Metz) acknowledges a fellowship from CNPq (Conselho Nacional de Desenvolvimento Cient\'ifico e Tecnol\'ogico), Brazil.

%%%%%%%%%%%%%%%%%%%%%%%%%%%%%%%%%%%%%%%%%%%%%%%%%%

%\section*{Appendix}

\appendix

\section{Stable distributions} \label{app:stable}

\begin{figure}[h!]
\begin{center}
\includegraphics[angle=-90, width=.6 \textwidth]{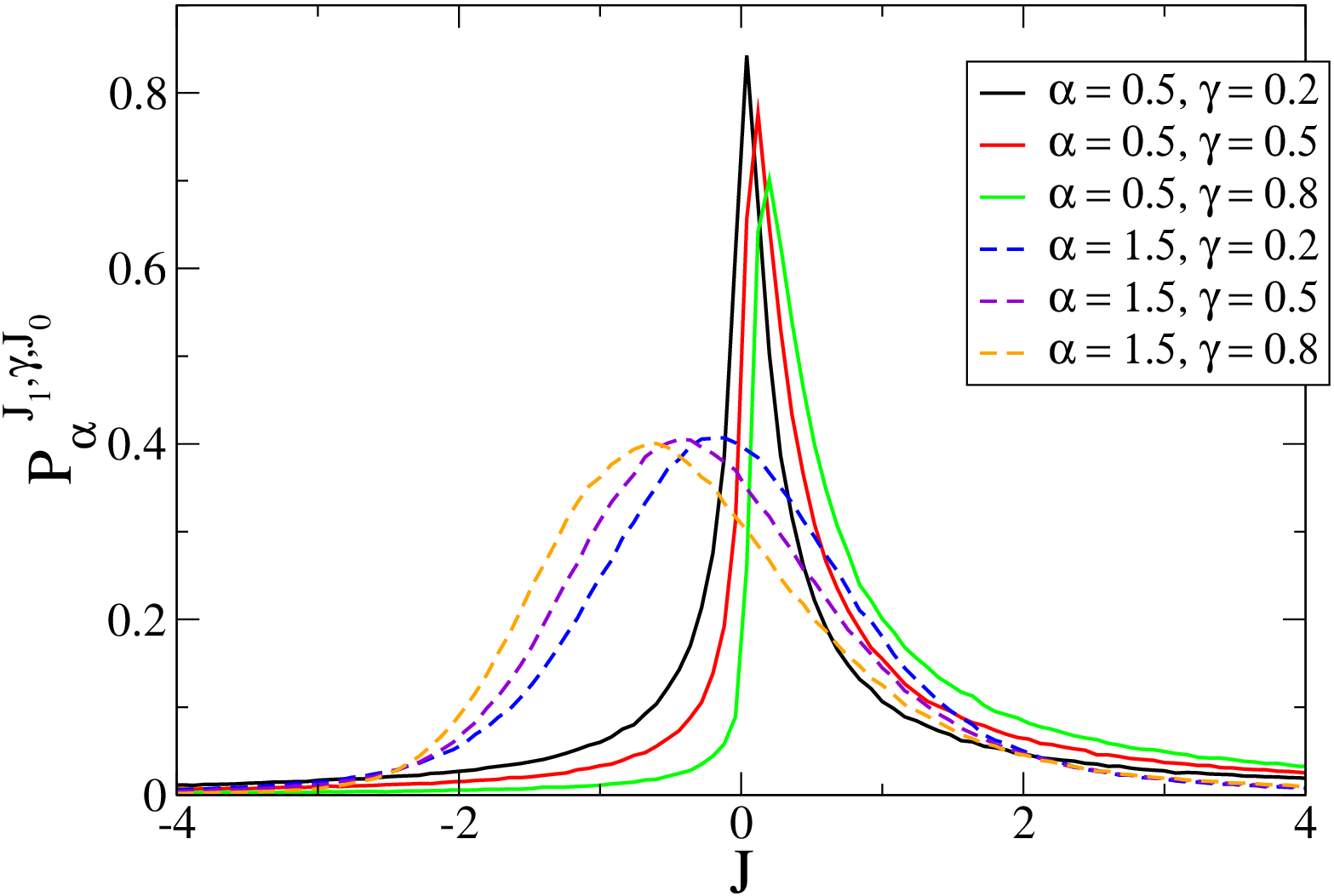}
\caption{The L\'evy distributions $P^{J_1, \gamma, J_0}_{\alpha}(J)$ for $J_1 = 1$, $J_0 = 0$ 
and different values of $\alpha$ and $\gamma$. The distributions with $\alpha=1.5$ 
approach a Gaussian while the ones for $\alpha=0.5$ have larger tails. 
%When $\alpha<1$ the 
%center of the distribution is positive for $\gamma>0$ while for $\alpha>1$ the center 
%is positive for $\gamma < 0$. 
For $\gamma > 0$, the center of the distribution
goes to $+\infty$ or $-\infty$ for $\alpha \uparrow 1$ or $\alpha \downarrow 1$, 
respectively. 
The coupling distribution fulfills $P^{J_1, \gamma, J_0}_{\alpha}(J)= 
P^{J_1, -\gamma, J_0}_{\alpha}(-J)$. }\label{fig:Levy}
\end{center}
\end{figure}
The purpose of this appendix is to give some
intuition on the role of the parameters $\alpha$ and $\gamma$ 
present in stable distributions, defined through 
eqs.~(\ref{defStable}) and (\ref{defChar}).
Both $\alpha$ and $\gamma$ are responsible for the
shape of the distribution. The main 
role of the exponent $\alpha$ is to control the decay of the tails.
Figure \ref{fig:Levy} shows that, for a fixed $\gamma$, a
decrease in $\alpha$ gives rise to a distribution 
$P^{J_1, \gamma, J_0 }_{\alpha}(J)$ with larger tails and 
more sharply peaked around its most probable value $J$. The
center of $P^{J_1, \gamma, J_0 }_{\alpha}(J)$ is also shifted
from the negative to the positive J-axis as $\alpha$ 
decreases from $\alpha > 1$ to $\alpha <1$. A change
of $\alpha$ has no effect on the position of the center
when $\gamma=0$. 

The skewness parameter $\gamma$
controls the relative weight between the positive and
negative tails. For 
$\gamma > 0$, the positive tail of $P^{J_1, \gamma, J_0 }_{\alpha}(J)$ is 
larger than the negative one; for $\gamma < 0$ vica versa.  The distribution is symmetric
around $J_0$ when $\gamma=0$. For increasing positive values
of $\gamma$, see figure 
\ref{fig:Levy}, the center of $P^{J_1, \gamma, J_0 }_{\alpha}(J)$ 
shifts to the right or left provided
$\alpha < 1$ or $\alpha >1$, respectively.
 
%%%%%%%%%%%%%%%%%%%%%%%%%%%%%%%%%%%%%%%%%%%%%%%%

\section{Solution of integrals} \label{app:int}

In this appendix we show how to integrate over $\hat{J}$ in the following 
equations:
\begin{eqnarray}
I_1 &=&
\int_{-\infty}^{\infty} \frac{dJ d\hat{J}}{2 \pi} |\hat{J}|^{\alpha} 
e^{i \hat{J} J } f(J), \label{defI1} \\
I_2 &=& 
\int_{-\infty}^{\infty} \frac{dJ d\hat{J}}{2 \pi} |\hat{J}|^{\alpha} 
{\rm sign}(\hat{J}) e^{i \hat{J} J } f(J), \label{defI2}
\end{eqnarray}
where $f(J)$ is given by eq.~(\ref{funcdefin})
and $\alpha \in (0,1) \cup (1,2]$.  We obtained the following 
results for $I_1$ and $I_2$ after integration over
$\hat{J}$
\begin{equation}
\fl
I_1= -\left(\frac{\sqrt{2}}{J_1}\right)^{\alpha} C_{\alpha} \int_{-\infty}^{\infty} \frac{dJ}{|J|^{\alpha+1}} 
\Big[ f(J)-f(0) \Big] , {\rm if} \quad 0 < \alpha < 2 , \label{finalI1}
\end{equation}
\begin{eqnarray}
\fl
I_2&=i \left(\frac{\sqrt{2}}{J_1}\right)^{\alpha} \frac{C_{\alpha}}{\Phi}
\int_{-\infty}^{\infty} \frac{dJ}{|J|^{\alpha+1}} {\rm sign}(J) f(J) 
 ,{\rm if} \quad 0 < \alpha < 1 , \label{finalI2A} \\
\fl &=i \left(\frac{\sqrt{2}}{J_1}\right)^{\alpha}\frac{C_{\alpha}}{\Phi}
\int_{-\infty}^{\infty} \frac{dJ}{|J|^{\alpha+1}}{\rm sign}(J) \Big[ f(J)-f^{\prime}(0)J \Big]
 ,{\rm if} \quad 1 < \alpha \leq 2 ,\label{finalI2B}
\end{eqnarray}
where the parameters $C_{\alpha}$ and $\Phi$ are defined 
in section \ref{sec:def}.  We have left out the 
dependence of $f(J)$ with respect to $h$ and $s$ since it is not 
important here.  The aim of this appendix is to show 
how one can derive eqs.~(\ref{finalI1}), (\ref{finalI2A}) and
(\ref{finalI2B}) from eqs.~(\ref{defI1}) and (\ref{defI2}).

By introducing an exponential convergence factor 
in eqs.~(\ref{defI1}) and (\ref{defI2}), we can
rewrite them as follows
\begin{eqnarray}
I_1 &=& 
\lim_{a \rightarrow 0^{+}} \int_{0}^{\infty} dJ \Big[f(J)+f(-J)\Big] \int_{0}^{\infty}\frac{d\hat{J}}{\pi} 
\hat{J}^{\alpha} \cos{(\hat{J} J)} e^{-a \hat{J}}  , \label{I1B}   \\
I_2 &=& i
\lim_{a \rightarrow 0^{+}} \int_{0}^{\infty} dJ \Big[ f(J)-f(-J) \Big] 
\int_{0}^{\infty}\frac{d\hat{J}}{\pi} \hat{J}^{\alpha} \sin{(\hat{J}J)} e^{-a \hat{J}}. \label{I2B}
\end{eqnarray}
The integrals over $\hat{J}$ are calculated for $a > 0$ 
and, afterwards, the limit $a \rightarrow 0$ is performed.
Reference \cite{Gradshteyn} can be used in
order to integrate over $\hat{J}$ in eqs.~(\ref{I1B}) and 
(\ref{I2B}), giving rise to 
\begin{eqnarray}
I_1 &=& \frac{\Gamma(\alpha+1)}{\pi}
\lim_{a \rightarrow 0^{+}} \int_{0}^{\infty} dJ \Big[ f(J)+f(-J) \Big] 
\frac{\cos{\Big[(\alpha+1){\rm arctan}(\frac{J}{a})\Big]}}{(J^2 + a^2)^{\frac{\alpha+1}{2}}}
  , \label{I1C}   \\
I_2 &=& i \frac{\Gamma(\alpha+1)}{\pi} 
\lim_{a \rightarrow 0^{+}} \int_{0}^{\infty} dJ \Big[ f(J)-f(-J) \Big] 
\frac{\sin{\Big[(\alpha+1){\rm arctan}(\frac{J}{a})\Big]}}{(J^2 + a^2)^{\frac{\alpha+1}{2}}}
 . \label{I2C}
\end{eqnarray}
In order to analyze the behavior of integrals (\ref{I1C}) and 
(\ref{I2C}) around $J=0$ when $a \rightarrow 0^{+}$, we insert a 
cutoff $\lambda > 0$ and split them as follows
\begin{eqnarray}
\fl
&I_1 &= \frac{\Gamma(\alpha+1)}{\pi}
 U_1(\lambda)  
+ \frac{\Gamma(\alpha+1)}{\pi}\cos{\Big[(\alpha+1)\frac{\pi}{2} \Big]}
\int_{\lambda}^{\infty} \frac{dJ}{J^{\alpha+1}} \Big[f(J)+f(-J)\Big] 
, \label{I1D}   \\
\fl
&I_2 &= i \frac{\Gamma(\alpha+1)}{\pi} 
U_2(\lambda)  + 
i \frac{\Gamma(\alpha+1)}{\pi}\sin{\Big[(\alpha+1)\frac{\pi}{2} \Big]}
\int_{\lambda}^{\infty} \frac{dJ}{J^{\alpha+1}} \Big[f(J)-f(-J)\Big]
 , \label{I2D}
\end{eqnarray}
where
\begin{eqnarray}
U_1(\lambda)= \lim_{a \rightarrow 0^{+}} \int_{0}^{\lambda} \frac{dJ}{a^{\alpha+1}} \Big[f(J)+f(-J)\Big] 
\frac{\cos{\Big[(\alpha+1){\rm arctan}(\frac{J}{a})\Big]}}
{\Big[ \Big(\frac{J}{a}\Big)^2 + 1\Big]^{\frac{\alpha+1}{2}}}, \label{U1A} \\
U_2(\lambda) = \lim_{a \rightarrow 0^{+}} \int_{0}^{\lambda} \frac{dJ}{a^{\alpha+1}} \Big[f(J)-f(-J)\Big] 
\frac{\sin{\Big[(\alpha+1){\rm arctan}(\frac{J}{a})\Big]}}
{\Big[ \Big(\frac{J}{a}\Big)^2 + 1\Big]^{\frac{\alpha+1}{2}}}. \label{U2A}
\end{eqnarray}
The limit $a \rightarrow 0^{+}$ has been 
performed on the right hand side of eqs.~(\ref{I1D}) and (\ref{I2D}).
The integrals present in the definition of $U_1(\lambda)$
and $U_2(\lambda)$ are computed 
through a power-series representation of their integrands, yielding
the results 
\begin{eqnarray}
\fl
U_1(\lambda) &=& \lim_{a \rightarrow 0^{+}} \sum_{n,l=0}^{\infty} u_{n l} \,
\Big( \frac{\lambda}{a} \Big)^{2l + \alpha +1} \lambda^{2 n - \alpha}  
, \label{U1}   \\
\fl
U_2(\lambda) &=& \lim_{a \rightarrow 0^{+}} \sum_{n,l=0}^{\infty} v_{n l} \,
\Big( \frac{\lambda}{a} \Big)^{2l + \alpha +2} \lambda^{2 n + 1 - \alpha}. \label{U2}
\end{eqnarray}
The explicit forms of the coefficients $\{ u_{n l} \}$ and 
$\{ v_{n l} \}$ are irrelevant. The analysis of 
eqs.~(\ref{I1D}) and (\ref{I2D}) as $\lambda$
tends to zero, constrained to the limit
$a \rightarrow 0^{+}$ in the functions 
$U_1(\lambda)$ and $U_2(\lambda)$, constitutes the
final step of the calculation.

One can notice from eq.~(\ref{U1}) 
that $U_1(\lambda)$ diverges for $\lambda \rightarrow 0^{+}$. 
However, the transformation of $f(J)$ according to $f(J) \rightarrow f(J) - f(0)$
removes this divergence and makes $U_1(\lambda)$ 
go to zero for $\lambda \rightarrow 0^{+}$, provided
that $\alpha < 2$. This allows us to perform
the limit $\lambda \rightarrow 0^{+}$ in eq.~(\ref{I1D})
which leads,
after comparison with eq.~(\ref{defI1}), to
the following result
\begin{eqnarray}
\fl
\int_{-\infty}^{\infty} \frac{dJ d\hat{J}}{2 \pi} |\hat{J}|^{\alpha} 
e^{i \hat{J} J } \Big[ f(J)-f(0)\Big] = -
\frac{\Gamma(\alpha+1)}{\pi}\sin{\Big(\frac{\alpha \pi}{2} \Big)} \nonumber \\
\times \int_{0}^{\infty} \frac{dJ}{J^{\alpha+1}} \Big[ f(J)+f(-J)-2f(0) \Big],\, 0 < \alpha < 2 . 
\end{eqnarray}
By integrating the term with $f(0)$ on the left hand
side of the above equation we get eq.~(\ref{finalI1}).

The calculation of eqs.~(\ref{finalI2A}) and (\ref{finalI2B}) proceeds in an
analogous way. Depending on the value of $\alpha$, there
are two different situations concerning the behavior
of eq.~(\ref{U2}) for $\lambda \rightarrow 0^{+}$.
For $\alpha < 1$, we obtain 
$\lim_{\lambda \rightarrow 0^{+}} U_2(\lambda)=0$, which
allows to perform the limit $\lambda \rightarrow 0^{+}$
in eq.~(\ref{I2D}). 
For $\alpha > 1$, 
it is necessary to transform $f(J)$ according to 
$f(J) \rightarrow f(J) - f^{\prime}(0) J$ in order
to obtain $\lim_{\lambda \rightarrow 0^{+}} U_2(\lambda)=0$ and to
perform the limit $\lambda \rightarrow 0^{+}$ 
in eq.~(\ref{I2D}).

\section*{References}
\bibliographystyle{ieeetr} % plain, alpha, ieeetr
\bibliography{bibliography}

\end{document}